# Chiral Walker breakdown in the U-shaped Permalloy nanotube

Zhenxuan He and Xinwei Dong*

Department of Physics and Institute of Theoretical Physics and Astrophysics, Xiamen University, Xiamen 361005, China

[Abstract] The motion of transverse domain walls (DWs) in U-shaped ferromagnetic nanotube which owns two different geometric confinements were investigated by micromagnetic simulation. Driven by unidirectional magnetic fields, the chirality-dependent DW structure and velocity under low fields and two phases of chiral Walker breakdown processes under high fields were observable, respectively. All these chirality-dependent behaviors can be attributed to the different dynamics of magnetizations in geometric confinements. Additionally, DW structures have different responses and sensitivity to the applied fields, leading to a hierarchy and complex Walker breakdown processes. This supplies a new perspective for manipulating DW chirality.

*To whom correspondence should be addressed.
Email address: dongxw@xmu.edu.cn

1. Introduction

The motion of domain walls (DWs) in ferromagnetic nanowires has been attracting great attentions for over a decade due to its promising applications in data storage [1] and logic gates [2]. The particle-like DWs can be propagated through complex networks of nanowires by applying magnetic fields [3-5] or electric currents [6-10]. When the velocity reaches a critical value, DW structures will change dynamically as they propagate, i.e. the famous Walker breakdown [11].

For the conventional case (i.e. in a flat strip), breakdown is mediated by the nucleation of a single antivortex at the lateral boundaries, followed with a periodic transformation between transverse and antivortex DWs, regardless of the chirality of DW [11]. More interestingly, in a cylindrical nanotube, a brand new process comes out with a synchronous motion of vortex-antivortex pair [12]. Recently, the manipulation of DW chirality is becoming much more meaningful due to its significant influence on DW behavior. Related research of DW chirality can be seen from recent reports [13-19]. Many systems with single geometric structure, like flat strip, have been deeply investigated. However, it should be mentioned that there is no obvious chirality-related difference between DWs in these reported systems. In this work, we designed a structure by combining two geometrical confinements. Consequently, we observed chirality-dependent DW structure and velocity, and achieved two phases of Walker breakdowns.

2. Micromagnetic simulation

We constructed a Permalloy nanotube with a U-shaped cross section, which simultaneously owns two kinds of geometric confinements: lateral boundary and arris. The diagrammatic sketch of U-shaped nanotube is observable in Fig.1 (a) and (b). The nanotube size is 6000×60×60 nm$^3$, with a thickness of 4 nm. It is consisted of three strips (*Strips* 1, 2 and 3), two arrises (*Arrises* 1 and 2) and two lateral boundaries (*Boundary* 1 and 2). For the convenience of observation, we unfold it into plane expansion when explaining the dynamic process of DWs in the following parts of the paper, as seen in the lower left of Fig. 1(a) and (b). We calculated the model by solving Landau-Lifshitz-Gilbert (LLG) equation with the Object Oriented Micro-Magnetic Framework (OOMMF). In our simulations, the related parameters of Permalloy are: saturation magnetization $M_s$ = 800 kA/m, exchange stiffness constant $A$ = 13 pJ/m, Gilbert damping constant $\alpha$ = 0.01 and zero crystalline anisotropy $K_1$=0. The mesh cell size is 4×4×4 nm$^3$.

The initial head-to-head DWs are set in the middle of nanotube to guarantee the stableness of original state. Like the case of flat nanostrip, there are also two kinds of DWs with different chiralities: counterclockwise (CCW) and clockwise (CW) DWs, determined by the initial magnetization arrangement in the cross section along DW center. The CCW or CW DW in the U-shaped nanotube is consisted of three DWs (i.e. DW1, DW2 and DW3 in *Strips* 1, 2 and 3, respectively) linked by the arrises, as shown in Fig. 1(c) and (e).

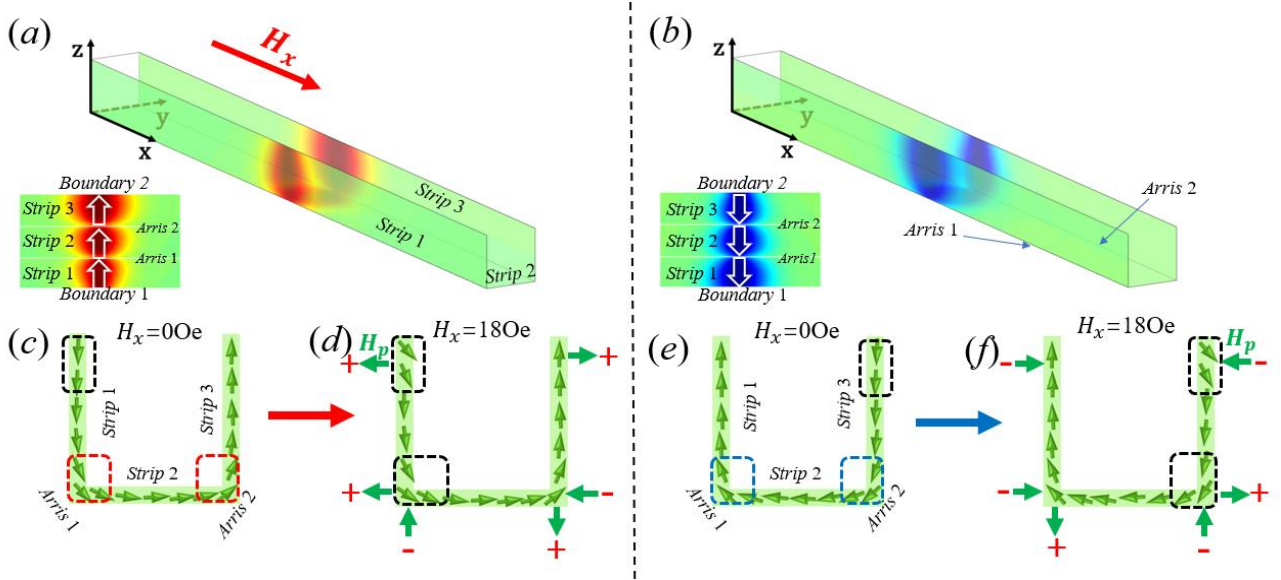

Fig. 1. The diagrammatic sketches and plane expansions for CCW (a) and CW (b) DWs. The white arrows represent the orientations of domains. The cross sections for CCW (c) and CW (e) DWs under $H_x$=0 Oe and for CCW (d) and CW (f) DWs under $H_x$=18 Oe.

3. Results and discussion

3.1 Chirality-dependent DW structure and velocity with fields below Walker field

To begin with, we firstly considered the initial state of DWs. Due to the distribution of demagnetization field ($H_{demag}$) and energy ($E_{demag}$), DW in each strip is strawberry-like shape [20] but with different DW widths, resulting into a pagoda-like shape as a whole. Considering that three DWs own different confinements (i.e. the lateral boundary and arris for DW1 and DW3, and two arrises for DW2), the magnetizations of arris ($M_A$) and lateral boundary ($M_{LB}$) both have influence on $H_{demag}$ of DW, especially for its perpendicular component ($H_p$). Note that $H_p$ in adjacent DWs induced by the same arris are mutually perpendicular. Thus, we define the positive value of $H_p$ as pointing outward while the negative represents for an inward direction (see Fig. 1(d) and (f)). Without field, $H_p$ in lateral boundaries are quite small while two orthogonal components of $H_p$ in adjacent DWs induced by each arris are numerically equal but with different orientations, as shown in Fig. 2(a) and (b). Thus, the initial CCW and CW DWs have nearly the same DW structures.

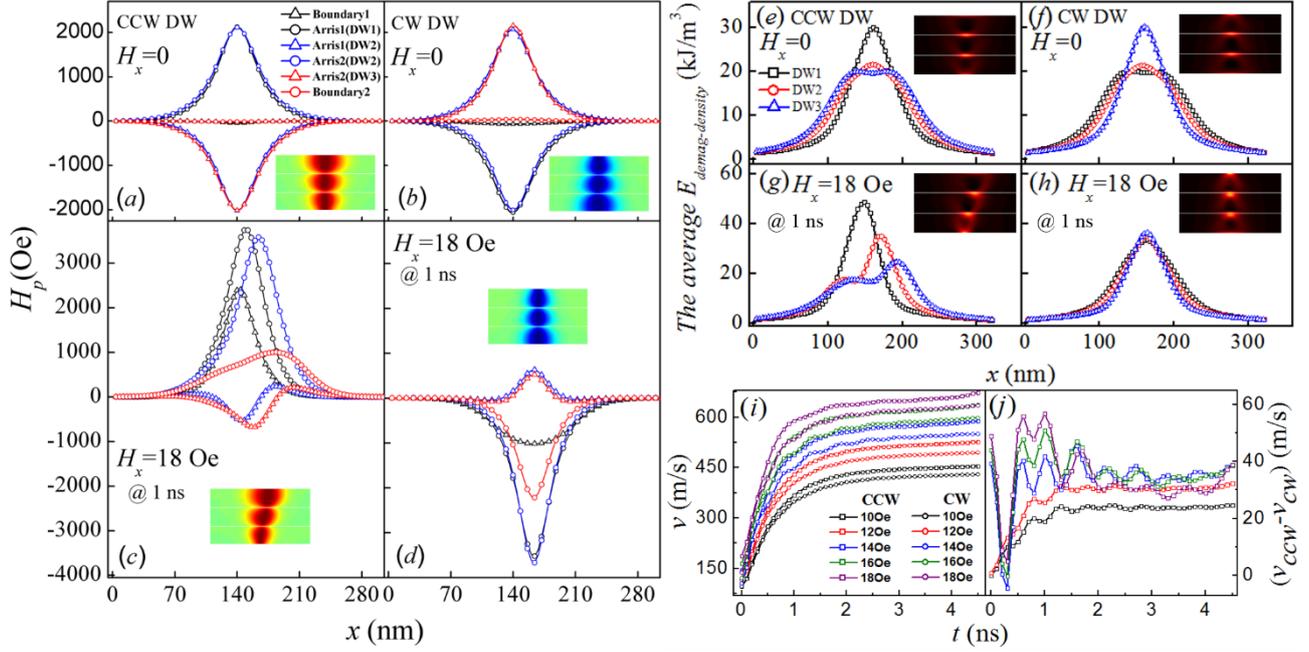

Fig. 2 The $H_p$ distributions in lateral boundaries and arrises along $x$-axis for CCW without field (a) and with field of 18 Oe (c), and for CW DW without field (b) and with field of 18 Oe (d), the insets show the plane expansions for DWs. The average $E_{demag\text{-}density}$ distributions along $x$-axis for CCW without field (e) and with field of 18 Oe (g), and for CW DW without field (f) and with field of 18 Oe (h), the insets show the detailed $E_{demag\text{-}density}$ distributions. The velocities of CCW and CW DWs (i) and their difference (j) under fields of 10 Oe, 12 Oe, 14 Oe, 16 Oe and 18 Oe.

When a field ($H_x$) below the so-called Walker field ($H_w$) is applied, magnetizations will rotate under the field torque. However, $M_A$ and $M_{LB}$ have different responses to the field, resulting into various change of DW widths in three DWs. In detail, for CCW DW under $H_x$ =18 Oe, the positive $H_p$ are increased while negative $H_p$ are decreased due to the rotation of $M_A$, together with the positive and newly-produced $H_p$ due to the rotation of $M_{LB}$, leading to the increment and decrement of $H_p$ in DW1 and DW3, respectively (see Fig. 2(c) and (g)). This means that the widest DW3 will get slight wider, and the narrowest DW1 will become narrower, resulting in the increment of width disparity among three DWs. The converse is true for CW DW, the widest DW1 will get narrower, and the narrowest DW3 will become wider, leading to the decrement of width disparity (see Fig. 2(d) and (h)). However, the DW width generally decreases with the increasing DW velocity. So, it is hard to distinguish the increment or decrement of each DW width, but their width disparity is observable all the time (see the insets of Fig. 2(a)-(d)). This can be further confirmed by the average $E_{demag\text{-}density}$ distributions of three DWs along $x$-axis, as shown in Fig. 2(e)-(h). Furthermore, the detailed $E_{demag\text{-}density}$ distributions are also observable in the insets of Fig. 2(e)-(h).

It is well known that wider DW always has a bigger velocity under the same field. To confirm that, we investigated the motion of CW and CCW DWs in a series of fields below $H_w$ (i.e. 10 Oe, 12

Oe, 14 Oe, 16 Oe and 18 Oe). As shown in Fig. 2(j), the velocity of stable CCW DWs under field of 10 and 12 Oe is ~24 m/s and ~31 m/s faster than that of CW DWs (with the time range of 2< $t$ <4.5 ns), respectively. The moving CCW and CW DWs both get unstable for the higher fields of 14 and 16 Oe. The velocity difference still oscillates in a large range of ~30-40 m/s even after 2 ns. When the field is increased up to 18 Oe, the instability of DWs is enhanced and the range of velocity difference is enlarged (~27-40 m/s). This implies that the chirality dependency of DW velocity gets weaker after a peak. As the field increases, Walker breakdown of CCW DW comes out at ~20 Oe, just slightly small than that of CW DW ($H_w$ =22 Oe).

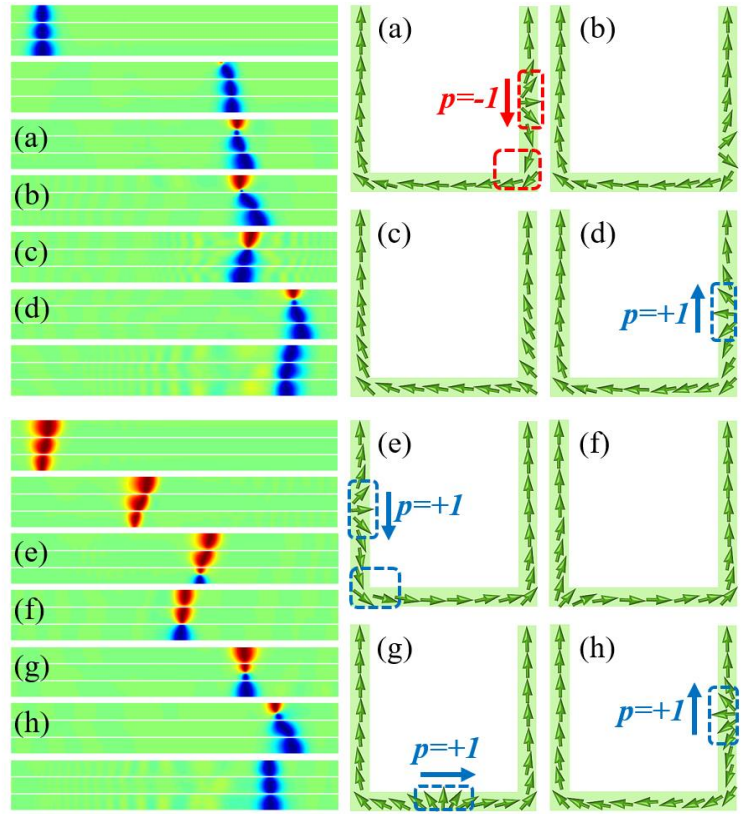

Fig. 3. On the left: the snapshots for the plane expansion of CW DW (upper part) and CCW DW (lower part) in the first-phase chiral Walker breakdown. On the right: the corresponding cross sections of (a)-(b) along DW center for CW DW and (e)-(h) for CCW DW, respectively.

3.2 The first-phase chiral Walker breakdown

Further, when a driving field slightly higher than $H_w$ is applied, the first phase of chirality-dependent Walker breakdown comes out. At beginning, an antivortex core (AVC) firstly forms at one lateral boundary and then propagates along the DW center (Fig. 2), but CW and CCW DWs behave very differently when their AVCs meet the arrises of nanotube. For CW DW, AVC bounces back after reaching *Arris*1 and then annihilates at the lateral boundary where it forms, leading to the maintenance of its CW-chirality (the upper part of Fig. 2). And CW DW keeps its own periodical procedure in the following propagation. However, for CCW DW, AVC can easily pass *Arrises* 1, 2 and annihilates at the other lateral boundary, changing its CCW-chirality into CW-chirality (the lower part of Fig. 2), and then repeats the periodic behaviors of CW DW.

Considering that there are two different polarities ($p$) of (anti)vortex in flat strip, i.e. $p =+ 1$ for upward orientation and $p = -1$ for downward orientation [21], we can accordingly define the polarity as $p = + 1$ for inward orientation (CCW DW) and $p = -1$ for outward orientation (CW DW) in the U-shaped nanotube. On the other hand, one can see that the perpendicular component of $M_{A1}$ (*Arris* 1) in *Strip* 1 has the same orientation with the $p$ of AVC in the case of CCW DW, but it is opposite in CW DW case (Fig. 3(a) and (e)). It also should be mentioned that $M_A$ is pinned due to the shape anisotropy of arris. For CCW DW, when AVC approaches *Arris* 1, the magnetizations between them are easy to be rotated (Fig. 2(f)), facilitating the propagation of AVC from *Strip* 1 to *Strip* 2 (Fig. 3(g) and (h)). But for CW DW, when AVC is approaching *Arris* 2, the intermediate magnetizations become frustrated, together with the strongly confined $M_{A2}$, finally changes the $p$ of AVC (Fig. 3(b) and (c)). It seems like that the oncoming AVC bounces back. The Walker breakdown in the U-shaped nanotube becomes chirality-dependent because the antivortex generated by $M_{LB}$, whose polarities are determined by chirality, correspond distinctively to the applied field.

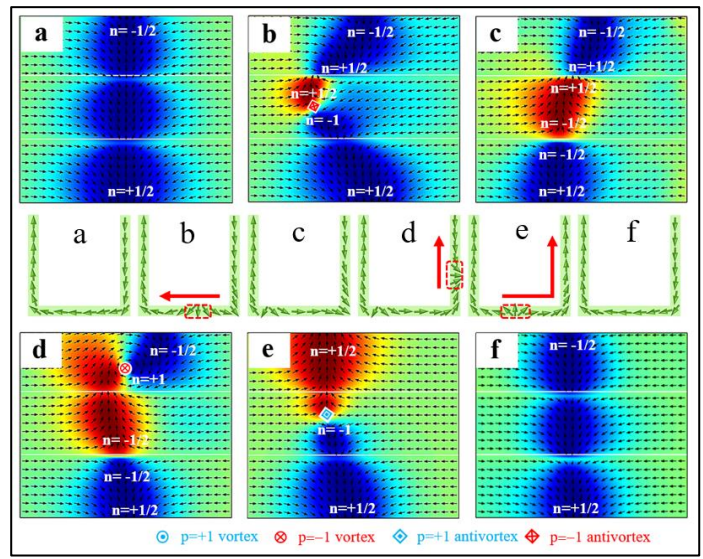

Fig. 4. The snapshots for the second-phase chiral Walker breakdown of CW DW. The winding numbers are tagged. (a) The original state of CW DW. (b) The generation of VC ($p=-1$) at *Arris*2 and AVC ($p=-1$) in *Strip* 2. (c) AVC moves to *Arris*1. (d) VC passes through *Strip* 3. (e) and (f) show the polarity-reversed AVC passing through *Strips* 2 and 3. (a)-(f) in the middle are the snapshots for the cross sections of CW DW center.

3.3 The second-phase chiral Walker breakdown

When the driving field is further increased, the chirality-dependent DW propagation gets into another different process. For a simple comparison, we name it as the second-phase Walker breakdown. For CW DW, when $H_x$ reaches a value of ~ 38 Oe, one vortex core (VC) is generated at *Arris* 2, followed with the nucleation of an AVC in *Strip* 2 to maintain the winding number (*n*) conservation of whole system (Fig. 4(b)). As following, a 90°-folded "VC" is trapped in *Arris* 2, but AVC in *Strip* 2 moves freely towards *Arris* 1, slowing down the velocity of DW2 (Fig. 4(c)). The enlarged velocity difference between DW2 and DW3 causes the depinning of "VC". Then, VC moves through *Strip* 3 and annihilates at *Boundary* 2, followed with the switch of DW3 chirality. Further, the polarity of AVC, which has been already blocked at *Arris* 1, is reversed by $M_A$ with similar mechanism mentioned in 3.2 (Fig. 4(d) and (e)). As expected, the polarity-reversed AVC smoothly passes through *Strips* 2 and 3. Consequently, CW DW turns back to the initial state after one period of propagation. In the following propagation, it can repeat the foregoing process (when $H_x$ reaches 44 Oe) or the first-phase process (when $H_x$ is below 44 Oe), depending on the magnitude of driving field.

For the second phase of CCW DW, when $H_x$ reaches a higher value of ~ 44 Oe, the displacement gap between DWs is enlarged due to the further increased velocity discrepancy, which has been mentioned in 3.1. For the narrowest DW1, its lateral boundary is potential to generate an AVC due to the high $E_{demag-density}$. However, the produced AVC cannot be injected because it cannot keep pace with the high-velocity DW1, although this process to some extent leads into the decrement of DW1 speed. Meanwhile, DW3 is accelerated all the time because its widest DW width protects it from being distorted. Thus, a larger displacement between DW2 and DW3 is created, leaving a 90°-folded 'AVC' in *Arris* 2 and a VC in *Strip* 2 (Fig. 5(b)). As following, the free VC continuously passes through *Strips* 2 and 1 under the gyrotropic force and annihilates at *Boundary* 1 (Fig. 5(b), (c) and (d)). Furthermore, the gyrotropic motion of VC definitely decreases the velocity of whole DW, which to some extent facilitates the injection of the AVC into *Strip*3. As expected, the released AVC from *Arris* 2 passes through *Strip* 3 and finally changes CCW DW into CW DW (Fig. 5(e) and (f)). In the following propagation, it will repeat the second-phase process of CW DW.

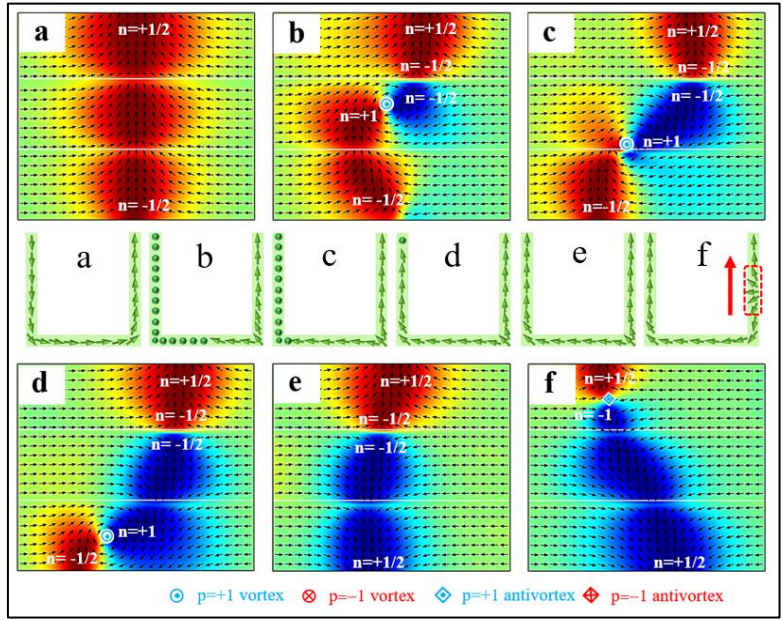

Fig. 5. The snapshots of the second-phase Walker breakdown for CCW DW. The winding numbers are tagged. (a) The original state of CCW DW. (b) The generation of AVC ($p$=+1) at *Arris*2 and VC ($p$=+1) in *Strip*2. (c)-(e) VC passes through *Strips* 1 and 2. (f) The depinning of folded AVC and the released AVC passes through *Strip*3. (a)-(f) in the middle are the snapshots for the cross sections along DW3 center. The dots in (b) and (c) represent that the magnetizations lie in *x*-axis.

The distinctions between CW and CCW in second-phase Walker breakdown primarily embody in different critical fields, which should be attributed to the different rotations of $M_A$. As well known that the antivortex or vortex DW consists of two +$^1/_2$ and one −1 or two −$^1/_2$ and one +1 topological defects, while $n$= +1 and $n$= −1 defects can be split in half to give the respective $n$= +$^1/_2$ and $n$= −$^1/_2$ defects on the edges of transverse DWs [22, 23]. For CW DW, the CW-chirality of DW3 determines the generation of a '90º-folded' VC at *Arris* 2, i.e. a head-to-head structure in *Arris* 2 (Fig. 4(b)). At meanwhile, the $p$ of VC (+1) is decided by the outward rotation of $M_{A2}$ under $H_x$. By contrast, for CCW-chirality of DW3 in the case of CCW DW, a '90º-folded' AVC with $p$ of −1 is produced at *Arris* 2, leaving a tail-to-tail structure in *Arris* 2 (Fig. 5(b)). According to the shape anisotropy of nanotube arris, the outward rotation of $M_{A2}$ is easier than the inward rotation (see Fig. 4(a) and (b), Fig. 5(a) and (b)). This can well explain why the critical field for CCW DW is ~6 Oe higher than that of CW DW. It is worth mentioning that if the applied field is further increased (>~48 Oe), the DWs will become unstable, followed with the disappearance of periodic propagation.

Additionally, different moving directions of the antivortex-vortex pair should be attributed to their different gyrotropic forces $\vec{F}_g = \vec{G} \times \vec{v}$, where $\vec{G}$ is the gyrovector, and $\vec{v}$ is in the direction of DW velocity. In detail, the gyrovector $\vec{G} = -2\pi pq\hat{n}$, where $p$ is the polarity of (anti)vortex, $q$ is the strength (+1 for vortex and −1 for antivortex), and $\hat{n}$ is the unit vector in the thickness direction, pointing towards the inner of nanotube [24,25]. Considering the same $p$ of antivortex-vortex pair, their gyrotropic forces are always in opposite directions under the same moving direction.

3.4 Conclusion

In conclusion, we constructed a Permalloy U-shaped nanotube, and achieved the chirality-dependent DW propagations in it. The velocity of DWs at low fields and Walker breakdown at high fields are all confirmed as chirality-dependent. Through the studies of physical mechanisms, the chirality-dependent behaviors can be attributed to the different micromagnetic structures generated from geometric confinements of U-shaped nanotube, together with their distinctive responses to applied fields, which take a significant role in the process and directly induce the chiral Walker breakdown. This supplies another way to control the DW chirality.

**ACKNOWLEDGEMENTS**

This work was supported by the National Natural Science Foundation of China (Grant No. 11204255) and Fundamental Research Funds for the Central Universities (Grant No. 20720160022).